\documentclass[preprint,showpacs,preprintnumbers,amsmath,amssymb]{revtex4}

\usepackage{graphicx}
\usepackage{dcolumn}
\usepackage[T1]{fontenc}
\usepackage[english]{babel}
\usepackage{amsmath}
\usepackage{amssymb}
\usepackage{tabularx}
\usepackage{multirow}
\usepackage[all]{xy}
\usepackage{appendix}

\begin{document}

\title{Higher-order Kerr terms allow ionization-free filamentation in gases}

\author{P. B\'ejot$^{1,2}$}
\author{J. Kasparian$^{1}$}\email{jerome.kasparian@unige.ch}
\author{S. Henin$^1$}
\author{V. Loriot$^{2}$}
\author{T. Vieillard$^{2}$}
\author{E. Hertz$^{2}$}
\author{O. Faucher$^{2}$}
\author{B. Lavorel$^{2}$}
\author{J.-P. Wolf$^{1}$}
\affiliation{(1) Universit\'e de Gen\`eve, GAP-Biophotonics, 20 rue
de l'Ecole de M\'edecine, 1211 Geneva 4, Switzerland}
\affiliation{(2) Laboratoire Interdisciplinaire CARNOT de Bourgogne (ICB), UMR 5209 CNRS-Universit\'e de Bourgogne, BP 47870, 21078 Dijon Cedex, France}

\begin{abstract}
We show that higher-order nonlinear indices ($n_4$, $n_6$, $n_8$, $n_{10}$) provide the main defocusing contribution
to self-channeling of ultrashort laser pulses in air and Argon at 800 nm, in contrast with the previously accepted mechanism of filamentation where plasma was considered as the dominant defocusing process. Their consideration allows to reproduce experimentally observed intensities and plasma densities in self-guided filaments.
\end{abstract}

\maketitle

The filamentation of ultrashort laser pulses in gases \cite{BraunKLDSM95} attracted a lot of interest in the last years because of its physical interest as well as its potential applications \cite{KasparianW08,BergeSNKW07,CouaironM07,ChinHLLTABKKS05}. Filaments are self-channeled structures propagating over many Rayleigh lengths without diffraction. They are generally considered to stem from a dynamic balance between Kerr focusing and defocusing by the plasma generated at the non-linear focus.
Numerical simulations based on this balance report a core intensity of several 10$^{13}$
W/cm$^{2}$ and typical electron densities of several 10$^{16}$ cm$^{-3}$ \cite{BergeSNKW07,CouaironM07}. Consequently, plasma ionization is generally admitted as necessary for an ultrashort pulse to experience self-channeling in gases.

But the plasma density provided by this description of filamentation appears overestimated as compared with experimental measurements. As reviewed in \cite{KasparianSC00}, such measurements are dispersed over several orders of magnitude, especially due to different focusing conditions and divergent assumptions about the core diameter of the filaments, but the electron density in a filament generated by a slightly focused beam is more likely to amount to $10^{14} - 10^{15}$ cm$^{-3}$ \cite{KasparianSC00}. This value, as well as the discrepancy by more than one order of magnitude with numerical simulations, was recently confirmed \cite{ThebergeLSBC06}. The observation of so-called plasma-free filamentation \cite{MechainCADFPTMS04, DubietisGTT04}, as well as the consideration that a balance between the instantaneous Kerr term and the time-integrated plasma contribution implies strongly asymmetric pulse shapes \cite{StibenzZS06}, periodically led to challenge the role of plasma in laser filamentation.

However, up to now, no other process seriously challenged plasma as the main defocusing process balancing the Kerr self-focusing. Nurhuda et al. proposed that the saturation of the nonlinear susceptibility $\chi^{(3)}$ should be taken into account \cite{NurhudaSM08}. Such saturation can be described as negative higher-order Kerr terms. The nonlinear index of air induced by high-power femtosecond laser pulses can be written as
$\Delta\text{n}_\text{Kerr}=n_2I+n_4I^2+n_6I^3+n_8I^4+ ...$
, where $I$ is the incident intensity and the $n_{2*j}$ coefficients are related to $\chi^\text{(2*j+1)}$ susceptibilities. This nonlinear index is generally truncated after its first term, $n_2$ \cite{KasparianW08,BergeSNKW07,CouaironM07,ChinHLLTABKKS05}, mostly because of the lack of data about the values of the subsequent terms.

Numerical works have investigated the influence of the quintic nonlinear response on the
propagation dynamics in gases, although without knowledge of its value
\cite{AkozbekSBC01,Couairon03,VincotteB04,FibichI04,Centurion05}. They showed that $n_4$ is negative,
\textit{i.e.} the $\chi^{(5)}$ susceptibility is a
defocusing term. It tends to stabilize the propagation of ultrashort laser
pulses in air and to decrease both the electron density and
the maximal on-axis intensity. Consequently, the losses due to
multiphoton absorption (MPA), which lead to the end of the filamentation,
are reduced and pulse self-channeling is sustained over longer
distances. However, plasma generation still appeared as necessary
for filament stabilization. Moreover, the value of $n_4$ was set arbitrarily,
which limits the conclusiveness of these studies. Finally,
the lack of data prevented any evaluation of a possible
effect of the further-order nonlinear refractive indices.

However, the higher-order Kerr indices have recently been measured in N$_2$, O$_2$ and Ar by Loriot \emph{et al.} \cite{Loriot09}. The reader is referred to this work for a detailed description of this experimental determination. In this Letter, we investigate their influence on numerical simulations of laser filamentation. We show that their values are sufficient to provide the dominant contribution to the defocusing terms of self-channeling. Their implementation in numerical simulations yields the experimentally observed plasma density. As a consequence, contrary to previously held beliefs, a plasma is not required for the observation of filamentation. Rather, plasma generation can be considered as a by-product of the self-guiding of laser filaments.

We implemented these nonlinear coefficients into a numerical model describing the propagation of ultrashort high power pulses
\cite{BejotBBW07}. We consider a linearly polarized incident electric field at $\lambda_0$=$800$ nm with cylindrical symmetry
around the propagation axis $z$. The scalar envelope $\varepsilon(r,t,z)$, assumed to vary slowly in time and along $z$, evolves according to the
propagation equation:

\begin{eqnarray}
\begin{aligned}
\label{Equation3}
&\partial_z\varepsilon =\frac{i}{2k_0}\triangle_{\bot}\varepsilon-i\frac{k''}{2}\partial_t^2\varepsilon+i\frac{k_0}{n_0}\left(\sum_{j=1}^{4}{n_{2*j}|\varepsilon|^{2*j}}\right)\varepsilon \\
&-i\frac{k_0}{2n_0^2\rho_c}\rho\varepsilon-\frac{\varepsilon}{2}\sum_{l=\mathrm{O}_2,\mathrm{N}_2}{\left(\sigma_l\rho+\frac{W_l(|\varepsilon|^2)U_l}{|\varepsilon|^2}(\rho_{at_l}-\rho)\right)}
\end{aligned}
\end{eqnarray}
where $k_0$=${2\pi n_0}/{\lambda_0}$ and $\omega_0={2\pi c}/{\lambda_0}$ are the
wavenumber and the angular frequency of the carrier wave respectively, $n_0$ is the linear refractive index at $\lambda_0$, $k''=\frac{\partial^2 k}{\partial\omega^2}|_{\omega_0}$ is the second order dispersion coefficient, $\rho_{at}$ the neutral atoms density, $\rho$ the electron density, $\rho_c=\epsilon_0 m \omega_0^2/e^2$ is the critical electron density, $m$ being the electron mass and $e$ its charge. $W_l(|\varepsilon|^2)$ and $\sigma_l$ are the photoionization probability and the
inverse Bremsstrahlung cross-section of species $l$ respectively (with ionization potential $U_l$), and $t$ refers to the retarded time in the reference frame of the pulse. The right-hand terms of Eq.(\ref{Equation3}) account for spatial diffraction, second order group-velocity dispersion (GVD), instantaneous nonlinear effects (\textit{i.e.} the nonlinear refractive index of air, up to the $n_8$ term), plasma defocusing, inverse Bremsstrahlung and multiphoton absorption respectively.
As compared with previously published data \cite{Loriot09}, we used values of the higher-order refractive indices (Table \ref{tab1}) incorporating the correction for the coherent artifact \cite{Oudar82}, \emph{i.e.} adequately substracting its electronic contribution at play in the original measurement of Ref. \cite{Loriot09}. This correction results in dividing each $n_{2*j}$ term by $j+1$.
Owing to the short pulse duration ($30\ fs$) used in the simulations, the delayed orientational response is disregarded. The propagation dynamics of the electric field is coupled with the density of the electrons originating from the ionization of both O$_2$ and N$_2$: $\rho=\rho_{\mathrm{O}_2}+\rho_{\mathrm{N}_2}$. This density is governed by the muti-species generalized Keldysh-PPT (Perelomov, Popov, Terent'ev) formulation \cite{KasparianSC00,BergeSNKW07}.

\begin{table}
\begin{tabular}{|c|c|c|c|c|}
\hline
& $n_2\ (10^\text{-19}$ & $n_4\ (10^\text{-33}$ & $n_6\ (10^\text{-46}$ & $n_8\ (10^\text{-59}$ \\
Species & $cm^2/W)$ & $cm^4/W^2)$ & $cm^6/W^3)$ & $cm^8/W^4)$ \\
\hline
N$_2$ & 1.1$\pm0.2$ & -0.5$\pm0.27$ & 1.4$\pm0.15$ & -0.44$\pm0.04$ \\

O$_2$ & 1.6$\pm0.35$ & -5.2$\pm0.5$ & 4.8$\pm0.5$  & -2.1$\pm0.14$ \\

Air & 1.2$\pm0.23$& -1.5$\pm0.3$& 2.1$\pm0.2$& -0.8$\pm0.06$ \\
\hline
\end{tabular}

\caption{Coefficients of the nonlinear refractive index
expansion of N$_2$ and O$_2$ at 1 bar pressure, and interpolation to air, as used in the present work \cite{Loriot09}} \label{tab1}
\end{table}

\begin{figure}
   \begin{center}
       \includegraphics[keepaspectratio, width=12cm]{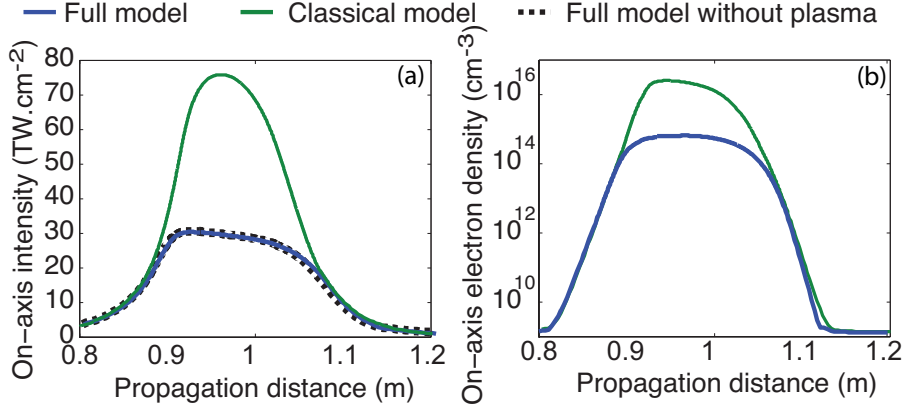}
   \end{center}
   \caption{(a) On-axis intensity and (b) Plasma density as a function of the propagation distance for the classical model (considering only $n_2$ term of the Kerr index and the plasma defocusing), the full model, as well as the full model without plasma.}
   \label{intensite_plasma}
\end{figure}

\begin{figure}
   \begin{center}
       \includegraphics[keepaspectratio, width=12cm]{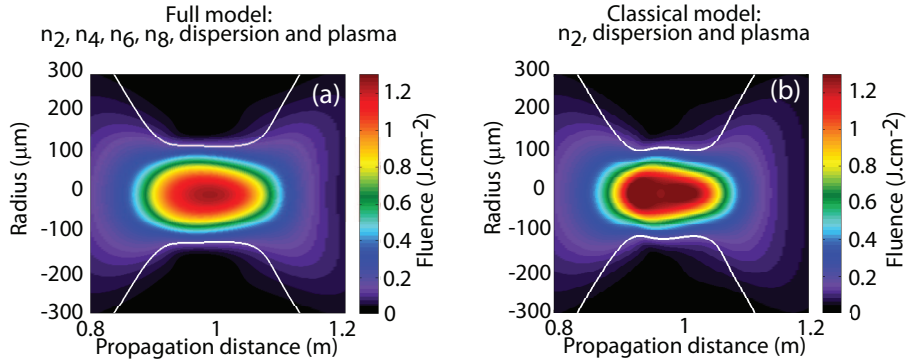}
   \end{center}
   \caption{Fluence distribution in air as a function of the propagation distance for the full model (a) and the classical model including $n_2$, ionization and GVD only (b). The white lines display the quadratic radius as a function of the propagation distance.}
   \label{intensite_z_r}
\end{figure}

We used this model to simulate the propagation of an ultrashort pulse typical of laboratory-scale experiments: $1$ mJ energy, $30$ fs FWHM pulse duration without initial chirp (hence, about 3.9 critical powers $P_{\mbox{cr}}$), an initial waist of $\sigma_{\mbox{r}} = 4$ mm, a focal length $f=1$ m and a pressure of $1$ bar.
Figures ~\ref{intensite_plasma} and \ref {intensite_z_r} compare the numerical results of the full model implementing Kerr
terms up to $n_8$ and of the classical model, where the Kerr term is truncated to $n_2$. Both models lead to self-guided filaments. The full model yields a lower maximum intensity ($31.6$ TW/cm$^{2}$ \textit{vs.} $78$ TW/cm$^{2}$), although these values lie within the range of published experimental data in comparable conditions \cite{BergeSNKW07,CouaironM07,ChinHLLTABKKS05,KasparianW08}. On the other hand, the full model predicts an electron density 40 times below the classical one ($1.1\cdot10^{15}$ cm$^{-3}$ \emph{vs.} $4.2\cdot10^{16}$ cm$^{-3}$). While the latter value is comparable with the output of other numerical works \cite{BergeSNKW07,CouaironM07,ChinHLLTABKKS05,KasparianW08}, the full model agrees with the available experimental measurements of the electron density \cite{KasparianSC00,ThebergeLSBC06}.

\begin{figure}
   \begin{center}
       \includegraphics[keepaspectratio,width=12cm]{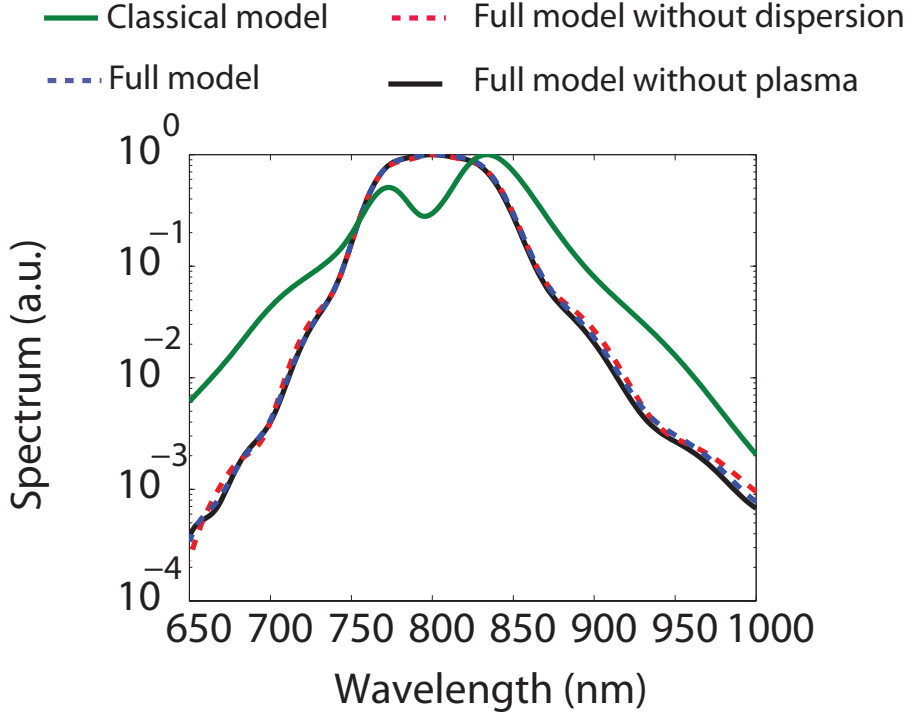}
   \end{center}
   \caption{Spectrum after 2 m propagation in air at atmospheric pressure.}
   \label{Spectre_2m}
\end{figure}

Note that, with the considered parameters, the full model yields a more strict intensity clamping than the classical one \cite{BeckerAVOBC01}. It predicts an intensity constant within $20\%$ over $15$ cm (\emph{vs} 9.5 cm in the case of the classical model), a length well comparable to experimental data reported to date in air for mJ-pulses \cite{BergeSNKW07,CouaironM07,HosseiniYLC04, MechainOFCPM06}.
This stricter clamping can be explained by the lower electron density, which results in weaker multiphoton losses, allowing a slower decay of the filament intensity and ionization.
The full model also yields a narrower output spectrum (Figure~\ref{Spectre_2m}), which better fits experimental data in air \cite{BergeSNKW07,CouaironM07}. It should therefore be considered as the reference model for numerical simulations of filamentation. Note that the almost symmetric shape of the spectrum is due to the neglection of self-steepening

On the other hand, neglecting the ionization in the full model (see Fig.~\ref{intensite_plasma}(a))  almost does not affect the simulation output. This shows that, in contrast to the classical understanding of filamentation in gases, the self-guiding process and plasma generation are almost decoupled. Instead, the negative higher-order nonlinear indices $n_4$ and $n_8$ constitute the dominant regularization terms leading to filamentation in air at atmospheric pressure. 
This limited influence of the ionization on the filamentation dynamics when higher-order non-linear indices are adequately considered sheds a new light on the possibility of ionization-free filamentation \cite{MechainCADFPTMS04}, which appears as a natural possibility in the context of the full model. Still, the dominant contribution of higher-order Kerr terms does not prevent ionization (Fig. \ref{intensite_plasma}b), which may contribute \emph{e.g.} to the conical emission.

\begin{table}
\begin{tabular}{|c|c|c|c|c|}
\hline
$n_2\ (10^\text{-19}$ & $n_4\ (10^\text{-33}$ & $n_6\ (10^\text{-45}$ & $n_8\ (10^\text{-59}$ & $n_{10}\ (10^\text{-74}$ \\
$cm^2/W)$ & $cm^4/W^2)$ & $cm^6/W^3)$ & $cm^8/W^4)$& $cm^{10}/W^5)$ \\
\hline
1.0$\pm0.09$ & -0.37$\pm1$ & 0.4$\pm0.05$ & -1.7$\pm0.1$ & 8.8$\pm0.5$  \\
\hline
\end{tabular}
\caption{Coefficients of the nonlinear refractive index expansion of Ar at 1 bar pressure, as used in the present work \cite{Loriot09}} \label{tab2}
\end{table}

\begin{figure}
   \begin{center}
       \includegraphics[keepaspectratio, width=12cm]{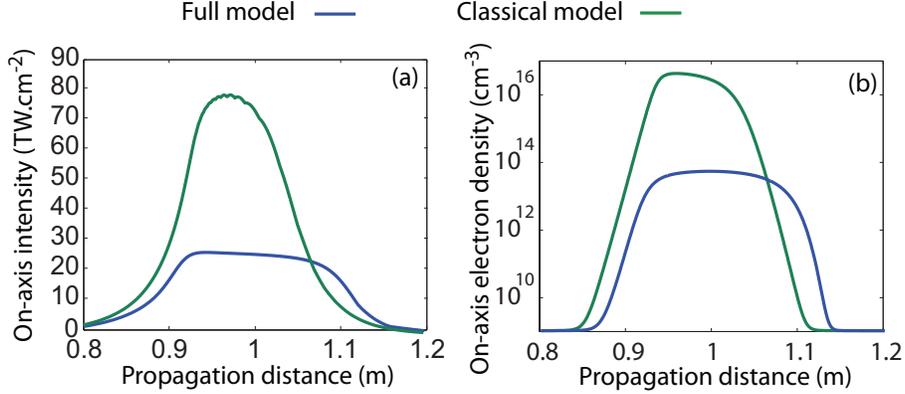}
   \end{center}
   \caption{(a) On-axis intensity and (b) Plasma density as a function of the propagation distance for the classical model (considering only $n_2$ term of the Kerr index and the plasma defocusing) and the full model, in Argon under 1 bar pressure.}
   \label{intensite_plasma_Ar}
\end{figure}

\begin{figure}
   \begin{center}
       \includegraphics[keepaspectratio, width=12cm]{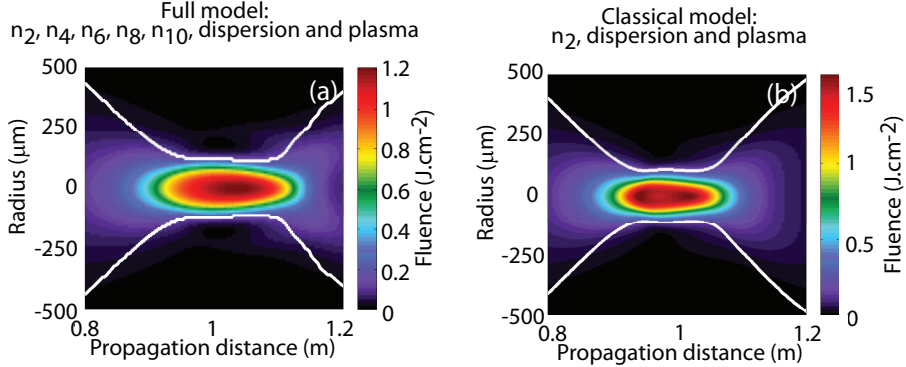}
   \end{center}
   \caption{Fluence distribution in 1 bar Argon as a function of the propagation distance for the full model (a), and the classical model including $n_2$, ionization and GVD only (b).}
   \label{intensite_z_r_Ar}
\end{figure}

We checked that the above conclusions are not restricted to a particular set of values of the non-linear refractive indices. Indeed, qualitatively comparable results have been obtained when varying the indices by several tens of percent, comparable with the experimental uncertainties on the non-linear indices. Furthermore, to compare the above molecular results with an atomic gas where no molecular orientation occurs, we performed simulations for Argon, where the ionization potential is close to that of the air molecules \cite{LoriotHLF08}, thus behaving in a similar manner as far as ionization is concerned. As in the case of air, we refined the corresponding indices to take the coherent artifact into account. The resulting values are summarized in Table \ref{tab2}. Like in air, the full model yields lower filament intensity ($28.5$ TW/cm$^{2}$ \textit{vs.} $80.9$ TW/cm$^{2}$) and electron density ($5.2\cdot10^{13}$ cm$^{-3}$ \emph{vs.} $4.1\cdot10^{16}$ cm$^{-3}$) than the classical model (Figure \ref{intensite_plasma_Ar}).
Also, the evolution of the fluence profile as a function of propagation distance  (Figure \ref{intensite_z_r_Ar}) is quite similar in both models.

\begin{figure}
   \begin{center}
       \includegraphics[keepaspectratio, width=12cm]{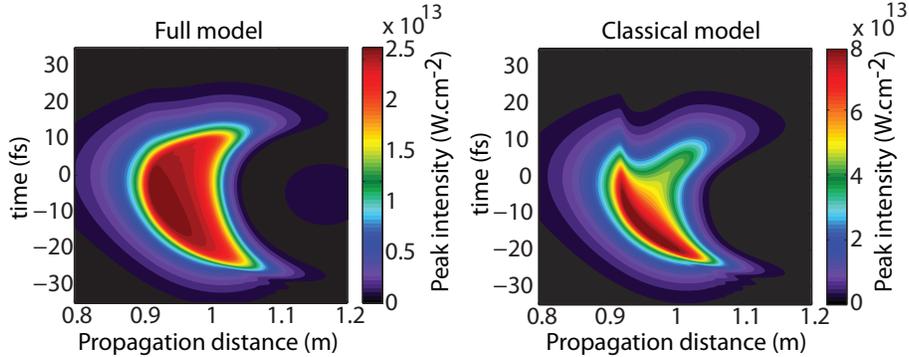}
   \end{center}
   \caption{Space-time dynamics of filamentation in 1 bar Argon for the full model (a), and the classical model including $n_2$, ionization and GVD only (b). Both models yield pulse splitting around 1.05 m propagation distance, but the full model where filamentation is driven by the instantaneous Kerr effect results in a more symmetrical temporal dynamics.}
   \label{intensite_z_t_Ar}
\end{figure}

The space-time dynamics shows more differences between the full and the classical models (Figure \ref{intensite_z_t_Ar}). In both cases, the pulse splits into two sub-pulses around 1.05 m propagation, but the full model predicts an almost symmetrical temporal profiles pattern all along propagation, while the classical model yields a largely asymmetric one. This behavior illustrates the different temporal dynamics of higher-order Kerr terms as compared with the plasma generation. The former is an instantaneous phenomenon depending only on the intensity. In contrast, the plasma generated during the pulse accumulates, resulting in an ever growing contribution. As a consequence, the leading edge of the pulse propagates in a low plasma density while the trailing edge is more defocused by the much higher electron concentration it encounters. Moreover, the lower losses due to the lower plasma density in the full model allows a slight refocusing cycle around 1.15 m, which is not predicted by the classical model. The results of the full model stay unaffected when the plasma is not taken into account (\emph{e.g} the peak intensity only increases by 0.6 \%), which confirms that the filamentation process, including the pulse splitting is indeed driven by the higher order Kerr terms when they are considered.

These differences in time-space dynamics illustrate the interest of implementing all orders of the Kerr effect in numerical simulations of filamentation in gases. Since successive terms $n_{2*j}I^j$ of the Kerr index are of alternate signs and have comparable values at an intensity of about $30-35$ TW/cm$^{2}$ \cite{Loriot09}, the inclusion of all terms up to $n_8$ in air (resp. $n_{10}$ in Argon) is necessary to adequately simulate the propagation of filamenting ultrashort pulses.

The observation that ionization, as well as GVD, almost do not affect the results of the full model provides an opportunity to speed up the numerical simulations. Neglecting the ionization typically cuts the computation time by a factor of 3 with little impact on the result in the conditions shown above. A parametric study would be necessary to determine the conditions, and especially the wavelengths and materials where such approximation is legitimate. Such study shall compare the intensities yielding a dynamic balance of the Kerr terms on one side, and between Kerr and plasma contributions on the other side. In air, where these intensities amount to $31.6$ TW/cm$^2$ and  $\sim78$ TW/cm$^2$, respectively, the lower intensity for pure Kerr balance ensures the domination of the latter process. Depending on the respective values of the higher-order non-linear indexes and ionization rates, the respective balance intensities may switch, leading to the domination of the Kerr-plasma balance.

In conclusion, we have shown that the recently measured higher-order nonlinear indices of air (up to $n_8$) or Argon (up to $n_{10}$) dominate both the focusing and defocusing terms implied in the self-guiding of ultrashort laser pulses in these gases. As a consequence, contrary to previously held beliefs, a plasma is not required to generate filamentation in gases, and its generation is quite decoupled from the self-guiding process. Instead, filamentation is, at least in the considered conditions, governed by higher-order nonlinear indices.
The usual definition of a filament as a dynamic balance between the $n_2$ Kerr self-focusing and defocusing on the plasma
shall therefore be revisited. Filamentation in gases rather appears as a nonlinear self-guided propagation regime sustained by a dynamic balance between nonlinear self-focusing and defocusing effects. Depending on experimental conditions, the latter can include higher-order Kerr terms and free electron with respective weights depending on the propagation medium considered.

Acknowledgements. This work was supported by the Conseil R\'egional
de Bourgogne, the ANR \textit{COMOC}, the \textit{FASTQUAST} ITN Program of the 7th FP, the Swiss NSF (contracts 200021-111688, 200021-116198 and R'equip program), as well as the SER in the framework of the Cost P18 program (Contract C06.0114)
\bibliographystyle{unsrt}

\end{document}